\documentclass[11pt]{article}

\usepackage{amsmath,amssymb,amsfonts}
\usepackage{epsfig}

\numberwithin{equation}{section}
\newcommand{\be}{\begin{equation}} \newcommand{\ee}{\end{equation}}
\newcommand{\bea}{\begin{eqnarray}} \newcommand{\eea}{\end{eqnarray}}
\title{
\begin{center}
 { \bf  On Witten's Instability and Winding Tachyons} 
\end{center}
}

\begin{document}

\author{Michael Dine, Assaf Shomer and Zheng Sun\footnote{\tt
dine,shomer@scipp.ucsc.edu,\ \ zsun@physics.ucsc.edu}}

\maketitle
\renewcommand{\thefootnote}{\fnsymbol{footnote}}

\centerline{\it   
Santa Cruz Institute for Particle Physics}
\centerline {\it 1156 High Street, Santa Cruz, 95064 CA, USA }

\begin{abstract}
We investigate, from a spacetime perspective,
some aspects of Horowitz's recent conjecture that
black strings may catalyze the decay of Kaluza-Klein spacetimes into a bubble of nothing.
We identify classical configurations that interpolate between
flat space and the bubble, and discuss the energetics
of the transition.  We investigate the effects of
winding tachyons on the size and shape of the
barrier and find no evidence at large compactification
radius that tachyons enhance the tunneling rate.
For the interesting radii, of order the string
scale, the question is difficult to answer
due to the failure of the $\alpha^\prime$ expansion.

\end{abstract}
\newpage
 
\setcounter{footnote}{0}
\renewcommand{\thefootnote}{\arabic{footnote}}

\section{Introduction}

The Kaluza-Klein vacuum,
empty $R^{1,d}\times S^1$, was shown many years ago \cite{Witten}\
to be non-perturbatively unstable.
The instability is against a process of semiclassical
barrier penetration that results in the formation of
a ``bubble of nothing" (BON) which expands at the speed of light
and ``eats up" the spacetime. 
This bounce solution possesses a number of intriguing features:
\begin{itemize}
\item It is admissible only when the boundary conditions on the circle
break supersymmetry.
\item
The solution is topology changing, so it is not
clear if and when it must be included in the path integral.
\item
Explicit solutions are known only for toroidal compactifications,
though one suspects that they may be generic
to non-supersymmetric
compactifications.
\item
From a four dimensional perspective, the radius of the fifth dimension is a modulus.
Quantum mechanically this modulus has a potential already
in perturbation theory.  If the potential
does not have a stable minimum, it is not clear
that the Witten solution is of any particular relevance.
If it does, it is not clear whether the instability persists.
\end{itemize}

The most common attitude towards this instability has been to ignore it, because the amplitude
for this process
\begin{equation}\label{rate}
 \Gamma = C e^{-{\pi R^2 \over 4  G_N}}
\end{equation} is exponentially small in the regime $R\gg \ell_p$ where the geometric description is reliable\footnote{Here $G_N\sim \ell_p^2$ is the four dimensional Newton's constant and $R$ is the radius of the KK direction.}. Speculative solutions to some of those problems have been raised e.g. in \cite{fabingerhorava}
and \cite{dfg}.

Consider a compactification of one of the
superstring theories on a six-dimensional torus, with all the radii
but one (associated with the direction $x^5$) of order the string scale.
Call $R$ the radius of the fifth dimension.  In order that a bounce solution
exist, we impose Scherk-Schwarz boundary conditions
in this direction.  The Witten bounce is a solution of the string equations
of motion to leading order in $\alpha^\prime$.  The stability analysis for the Schwarzschild-Tangherlini black hole \cite{regge} \cite{vishvesh} can be carried over to this case (see discussion in \cite{ofer}\ and \cite{gibhart}\ section 4.4) and implies that there are no zero modes\footnote{By this we mean that there are no non-trivial zero modes. Of course, there are zero modes that correspond to the rotation and translation symmetries of these solutions.} for the quadratic fluctuation operator. Therefore, one expects to be able to compute corrections to the solution systematically for large $R$ in powers of $\alpha^\prime/R^2$.
In terms of the string coupling, $g_s$, and length
scale, $\ell_s$, the leading order action, eqn. \ref{rate}, takes the form:
\begin{equation}
 S^{(0)}_{w} =  {A \over g_s^2} \bigl( {R\over \ell_s}\bigr)^3
\end{equation} 
where $A$ is a constant of order unity.  The solution and the action
will receive corrections in powers of $\alpha^\prime$
\begin{equation}
 S_w = S^{(0)}_{w}\cdot f({R\over \ell_s}) 
\end{equation} 
where $f \rightarrow 1$ as $R \rightarrow \infty$.
One can imagine various possible behaviors for $f$ as $R \rightarrow \ell_s$.
If it is a positive number of order one, then the tunneling rate
remains suppressed at weak coupling.  However,
if the function $f$ vanishes or becomes very small,
the rate for tunneling becomes order one,
corresponding to the disappearance (or near disappearance)
of the barrier between flat space and the bubble configuration.

In low orders of the $\alpha^\prime$ expansion for the solutions,
there is no clear signal for what the behavior might be at $R\sim \ell_s$.
But it has long been observed that within string theory the existence of 
Witten's instability is often associated
with the presence of winding tachyons
(see for example \cite{barbon} \cite{minwalla} \cite{gutperle}).
Indeed, Witten's solution is only admissible if
the fermions obey anti-periodic boundary conditions around
the KK circle.  For such boundary
conditions, the Type II and the heterotic
string theories develop a perturbative tachyon
instability \cite{rohm}\ for a sufficiently small radius.
It is possible that the appearance of tachyons will tend to decrease $f$.

\subsection{Horowitz's conjecture}

Recently, Horowitz \cite{horowitz} has argued that black strings can catalyze Witten's decay process.
Horowitz starts with the observation
that in the presence of the black string,
the compactification radius (for toroidal
compactifications) is a function of position.  For suitable choice
of charges, the radius can be arbitrarily slowly varying,
and can approach the string scale near the horizon, even
if the radius is large at infinity.
In this region, a tachyon appears. Horowitz
observes an analogy with the work of ref. \cite{pinch}\ where
in a similar situation the appearance of a winding tachyon
caused a topology changing
transition where a circle pinched off the original geometry.
The geometry in this case
is similar to that
of Witten's bubble solution
(which will be reviewed in detail in the next section),
where the radius of the KK direction shrinks to zero size as one approaches the bubble wall.
The work of
\cite{pinch}
was based on the correspondence of exponentially growing
tachyon field configurations with certain Liouville conformal field theories.
Horowitz notes that this could represent a previously unsuspected endpoint for Hawking radiation.  
This is an intriguing observation, but the physics is in many ways
obscure. For instance, such world-sheet theories do not include some of the features
(e.g. the various moduli and approximate moduli)
one typically finds in critical string models.

In this note we try and gain some understanding
of Horowitz's suggestion from a
{\it spacetime} perspective. The basic observation we use is that for this analysis to make sense the geometry in the vicinity of the horizon is assumed to be nearly flat. This is necessary for there to even be a winding tachyon a-la Rohm \cite{rohm}. Therefore, {\it the vicinity of the horizon is itself subject to Witten's flat-space instability.}
There is however a physical difference. Witten's instability is an
instability in pure GR. Here, the presence of the black string ``brings down" a scalar field that becomes tachyonic near the horizon
(a {\it quasi-localized} tachyon, in the terminology of \cite{quasi}).
Applying one's
naive field theoretic intuition, Horowitz's claim may be understood as the statement that what used
to be a {\it non-perturbative} instability (a tunneling event)
in the empty KK vacuum becomes a {\it perturbative}
instability (shrinking the
potential barrier or rolling down a tachyonic potential) in the presence of the black string.

To develop a space-time
picture, we first examine the
potential barrier for the Witten process.
We then proceed to ask whether Witten's {\it flat space}
instability is enhanced due to the presence of a tachyon at small radius.
In the regime where
we can trust the $\alpha^\prime$ expansion,
the answer to this last question is sensitive
to order one, model dependent numbers.
In type II string theory, there
does not appear to be any
enhancement, but the tachyon does exhibit a zero mode in the BON background. We speculate on possible
behaviors once higher order correction in $\alpha^\prime$ are included.
Recent related works include, for
example, refs. \cite{ross} \cite{oren}\ \cite{nakayama}.

\section{Potential Barrier for the Witten Bounce}

In standard analysis of vacuum decay, one starts with a theory with a known false and true
vacuum, and looks for solutions which in the far past (in Euclidean time) asymptote to
a bubble of true vacuum, large enough that it is energetically favored for it to
grow, and to the false vacuum in the far future\cite{coleman}.  This can be understood
as a conventional WKB calculation of quantum mechanical tunneling over
a barrier\cite{sakita,bbw}.  In a field theory, one can actually consider many trajectories;
the {\it bounce solution} can then be thought of as the most probable tunneling
path, others giving a lower bound on the tunneling rate.  In theories
coupled to gravity, the situation is more subtle, but the picture is similar\cite{cdl,bankscdl}.

Witten obtained his bounce in an approach which seems more abstract.  He exhibited
a solution of the Euclidean five dimensional Einstein equations which had finite
action, and a single negative mode.  To interpret the solution he followed the prescription of \cite{cdl}, analytically continuing the solution to Minkowski signature and
demonstrating that it corresponded to a decay to a bubble which grows.  The bubble
has the bizarre feature that, from a four dimensional perspective, it has no interior;
space time ends at the bubble wall\footnote{Another question is the nature of the true vacuum ``nothing" state.
Within string theory the disappearance of the spacetime manifold
itself which can be loosely thought of as the ``arena" for closed
string modes seem to resonate well with processes in
the open string sector where as a result of an instability the
``arena" for open strings, namely D-branes, disappear.
We do not offer here any new insights into this problem.}.

In this section we will construct a set of {\it classical configurations}
of the theory that smoothly interpolate between the KK vacuum and the Witten
bubble at $t=0$. 
We will require that these configurations represent suitable initial data
in general relativity, so that it makes sense to think of their energy as potential
energy.  The ADM energies of this sequence of configurations then describe a potential
barrier.  The sequence of configurations is a possible tunneling trajectory.
Motion over this barrier, in the sense of WKB, gives a tunneling trajectory
to Witten's bubble configuration, though not necessarily the lowest action trajectory.
As we will see, the ADM energies of these static solutions indeed give a potential barrier with the expected properties.

We start by briefly describing Witten's bounce.
Begin from the metric describing the KK vacuum in $D=n+3$ dimensions
\begin{equation}
ds^2=-dt^2+dx_1^2+\dots+dx_{n+1}^2+d\chi^2
\end{equation} 
with $t,x_i$ running from $-\infty$ to $\infty$ and $\chi\sim\chi+2\pi R$.
We now analytically continue to Euclidean signature
\begin{equation}
ds^2_E=dt_E^2+dx_1^2+\dots+dx_{n+1}^2+d\chi^2
\end{equation} 
and change to spherical coordinates by defining $r^2=t_E^2+\vec{\bf x}\cdot\vec{\bf x}$
\begin{equation}
ds^2_E=dr^2+r^2 d\Omega_{n+1}^2+d\chi^2
\end{equation}
The Euclidean bounce solution mediating the decay into a BON is an analytical
continuation of the $D-$dimensional Schwarzschild solution 
\begin{equation}\label{ebounce}
ds^2=\dfrac{dr^2}{1-\bigl( \frac{Rn/2}{r}\bigr)^n}
+r^2d\Omega_{n+1}^2+\Bigl(1-\bigl(\frac{Rn/2}{r} \bigr)^n\Bigr)d\chi^2
\end{equation} where regularity requires that we set the ``horizon" radius to be at
$\frac{Rn}{2}$ and restrict the radial coordinate $r\geq \frac{Rn}{2}$.

For the Lorentzian bounce we write $d\Omega_{n+1}^2=d\theta^2+\sin^2\theta d\Omega_n^2$ and analytically continue $\theta\rightarrow \frac{\pi}{2}+iT$ to get
\begin{equation}\label{lbounce}
ds^2=\dfrac{dr^2}{1-\bigl(\frac{Rn/2}{r} \bigr)^n}+r^2\bigl(-dT^2+\cosh^2 Td\Omega_n^2\bigr)
+\Bigl(1-\bigl(\frac{Rn/2}{r} \bigr)^n\Bigr)d\chi^2
\end{equation} which looks to an observer in $n+2$ dimensions (assuming $R$ is small and that one is not probing too close to $r=\frac{Rn}{2}$) as if a hole of radius $\frac{Rn}{2}$ appeared at time $T=0$ in Minkowski space and then keeps expanding until it ``eats up" all of space.

Another known bubble solution in the same $R^{1,n+1}\times S^1_R$, the {\it static bubble}
\cite{Sarbach:2004rm}\cite{Corley:1994mc},
is a product of the $n+2-$dimensional Euclidean Schwarzschild solution and a
trivial time direction
\begin{equation}\label{static}
ds^2=-dT^2+\dfrac{dr^2}{1-\bigl(\frac{R(n-1)/2}{r} \bigr)^{n-1}}
+r^2d\Omega_n^2+\Bigl(1- \bigl(\frac{R(n-1)/2}{r} \bigr)^{n-1}\Bigr)d\chi^2.
\end{equation} From the perspective of an $n+2-$dimensional observer this is a static massive BON whose ADM mass is given by
\begin{equation}\label{staticmass}
\mathcal{M}_{n+2}=\dfrac{\Omega_n}{16\pi G_{n+2}}\bigl(\dfrac{R}{2}(n-1)\bigr)^{n-1}.
\end{equation}  
Notice that the size of this bubble $\frac{R(n-1)}{2}$ is smaller than the size of Witten's bubble at the moment of formation  $\frac{Rn}{2}$.
In fact, this is very natural
from the usual perspective of vacuum decay.
The dynamics of true vacuum bubbles are determined by a competition between the negative energy contribution of the bubble (by definition the true vacuum has lower energy than the false vacuum) and the positive energy due to the tension of the bubble wall. A small true vacuum bubble will re-collapse while those larger than a critical size will keep expanding. At the critical size there is always a static unstable bubble. Indeed \ref{static}\ is known to be classically unstable, e.g. by relating it via double analytical continuation to the Gregory-Laflamme instability of black strings \cite{Sarbach:2004rm}. Also, Witten's bounce \ref{lbounce}\ is massless from the perspective of this $n+2$-dimensional observer since it should be thought of as a decay of the vacuum. The positive {\it kinetic} energy of the bubble is balanced against a negative energy associated with the opening of the hole in space.

We now demonstrate how to build initial data corresponding to BON of arbitrary size.
This will lead directly
to the construction of the potential barrier for Kaluza-Klein decay.
Let us first examine the Cauchy data for \ref{lbounce}\ and \ref{static}\ at
an initial time slice that we take to be $T=0$. The Cauchy data for \ref{lbounce}\
is given by
\begin{equation}\label{lbouncedata}
ds_{Cd}^2=\dfrac{dr^2}{1-\bigl(\frac{Rn/2}{r} \bigr)^n}+r^2d\Omega_n^2
+\Bigl(1-\bigl(\frac{Rn/2}{r} \bigr)^n\Bigr)d\chi^2.
\end{equation} while for \ref{static}\ it is given by
\begin{equation}\label{staticdata}
ds_{Cd}^2=\dfrac{dr^2}{1-\bigl(\frac{R(n-1)/2}{r} \bigr)^{n-1}}
+r^2d\Omega_n^2+\Bigl(1-\bigl(\frac{R(n-1)/2}{r} \bigr)^{n-1}\Bigr)d\chi^2.
\end{equation}
Assuming spherical symmetry we are led to look at Cauchy data of the general form
\begin{equation}\label{anzats}
ds_{Cd}^2=\dfrac{dr^2}{f_n(r)}+r^2d\Omega_n^2+f_n(r)d\chi^2.
\end{equation} 
We now need to impose the initial data constraint (see e.g. \cite{wald})
\begin{equation}
G_{\mu 0}=T_{\mu 0}.
\end{equation} 
Since this is a vacuum solution (namely just the dynamics of gravity without matter) we can set $T_{\mu\nu}=0$. Furthermore, since we are looking for a time symmetric (bounce) solution we can choose the initial data {\it at the turning point} where time derivatives of the metric vanish so $^{(n+3)}\mathcal{R}_{\mu 0}=0$. This simplifies the constraint to a vanishing of the $n+2-$dimensional scalar curvature
\begin{equation}
^{(n+2)}\mathcal{R}=-\frac{1}{r^2}\bigl[n(n-1)\bigl(f_n(r)-1\bigr)+2nrf_n^{\prime}(r)+r^2f_n^{\prime\prime}(r)\bigr]=0
\end{equation} 
which is easily solved to give (for $n\geq 2$, namely $D\geq 5$)
\begin{equation}
f_n(r)=1-\dfrac{a_n^{n-1}}{r^{n-1}}+\dfrac{b_n^n}{r^n}
\end{equation} with $a_n,b_n$ arbitrary constants.

The parameter $a_n$ is simply related to the $n+2-$ dimensional ADM energy of
the solution\footnote{Notice that here we mean the energy as measure by an observer in the $n+2$-dimensional space spanned by $t,r,\Omega_n$ one gets after KK reduction on $\chi$. This should not be confused with the $n+2-$dimensional space like Cauchy surface.}
\begin{equation}\label{adme}
\mathcal{M}_{n+2}=\dfrac{\Omega_n}{16\pi G_{n+2}}a_n^{n-1}
\end{equation} where $\Omega_n$ is the volume of the unit n-sphere $S^n.$
This Cauchy data will in general be singular where $f(r)=0$. Let us
denote by $r_0$ the largest real solution and demand that the Cauchy data will be regular by the usual absence of a conical singularity near $r=r_0$ 
\begin{equation}\label{regcon}
\dfrac{4\pi}{f_n^{\prime}(r_0)}=2\pi R.
\end{equation}  
Even though we can not generally (for $n>2$) solve for $r_0$ we can bypass this technicality by using $r_0$ as our variable. In fact, this will give a bubble solution where $r_0$ is the position of the wall so it is the natural variable to use.
Writing down the equation
\begin{equation}\label{arnot}
f_n(r_0)=1-\dfrac{a_n^{n-1}}{r_0^{n-1}}+\dfrac{b_n^n}{r_0^n}=0
\end{equation} 
and plugging that into \ref{regcon}\ we obtain the following relation between the energy of the solution and the radius of the bubble wall at $T=0$
\begin{equation}\label{aeiaen}
f_n^{\prime}(r_0)=\dfrac{1}{r_0}[n-(\dfrac{a}{r_0})^{n-1}]=\dfrac{2}{R}
\end{equation} 
which via \ref{adme}\ gives the desired relation between the energy and the radius of the bubble at formation
\begin{equation}\label{pot}
V(r_0)=\dfrac{\Omega_n}{16\pi G_{n+2}}\dfrac{r_0^{n-1}}{R}(nR-2r_0)
\end{equation} 
Defining for brevity 
\begin{equation}
r_{W}=\dfrac{Rn}{2} \qquad \rm{and}\qquad  r_{S}=\dfrac{R(n-1)}{2}
\end{equation} to be the corresponding sizes of Witten's bubble and the static bubble at $T=0$ we can rewrite this as
\begin{equation}\label{bpot}
\boxed{V(r_0)=\dfrac{\Omega_n}{8\pi G_{n+2}}\dfrac{r_0^{n-1}}{R}(r_W-r_0).}
\end{equation} 
In all dimensions $n>2\ \ (D>5)$ this potential looks\footnote{For the special case of $n=2\ \ (D=5)$ the picture looks less smooth at the origin.} like figure 1.
\begin{figure}\label{figone}
 \centering
 \includegraphics[width=7cm,height=5cm]{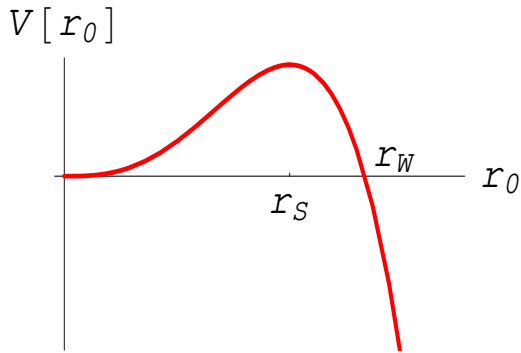}
\caption{{\bf The shape of the potential barrier penetrated by Witten's instanton.}\newline This figure plots the potential as a function of $r_0$ which is the size of the bubble at the moment it materializes. There are two zero energy solutions corresponding to the KK vacuum at $r_0=0$ and Witten's bubble at $r_0=r_W$. The unstable static bubble is at $r_0=r_S$. The height of the potential at the top is of order $V(r_S)\sim (\frac{R}{\ell_p})^{n-1}\cdot m_p$. }
 \label{fig1}
\end{figure}

We can now verify that this is the expected barrier. The potential vanishes in two places.
One is when $r_0=0$ which is the KK vacuum.
The other is when $r_0=\frac{Rn}{2}$ which is the radius of Witten's bounce
in $n+3$ dimensions; at this point, the configuration
is precisely the bubble at $T=0$ (see equation \ref{lbounce}).
The extrema of \ref{bpot}\ are a local minimum (only for $n>2$)
at $r_0=0$ which corresponds to the perturbative stability of
the KK vacuum and a local maximum at $r_0=\frac{R(n-1)}{2}$, at which
point the configuration is the perturbatively unstable static bubble \ref{static}.

Using \ref{arnot}\ and \ref{aeiaen}\ we can rewrite $f_n(r)$ describing the initial data \ref{anzats}\ in terms of $r_0$
\begin{equation}
f_n(r)=1+\dfrac{2}{R}(\dfrac{r_0}{r})^{n-1}(r_0-r_{W})-\dfrac{2}{R}(\dfrac{r_0}{r})^n(r_0-r_{S}).
\end{equation} 
The exact (time dependent) solutions are known only for 3 points along this curve, corresponding to the KK vacuum $(r_0=0)$, the static bubble at the top of the hill $(r_0=r_S)$ and Witten's bounce \ref{lbouncedata}\ $(r_0=r_W)$. 
For the case $n=2$ this initial data is a one dimensional subset of a two parameter family found by Horowitz and Brill \cite{Brill:1991qe}\ who used it to demonstrate the existence of solutions with arbitrarily negative energy in GR. The physical picture presented here is also supported by the analysis of \cite{Corley:1994mc}. 

We end this analysis by stressing that the path in field space indicated by the single
degree of freedom $r_0$ in figure 1. does not represent the actual (minimal action)
path in field space traversed by Witten's instanton as it tunnels through the barrier.
If one were to write the Witten solution in a gauge with $g_{tt}=1$, $g_{\mu t}=0$
(singling out one coordinate to call the Euclidean time), at general points along the
curve the time derivatives (and the kinetic energy) will be non-zero.
Still, this potential barrier captures
the qualitative features of the actual path and demonstrates that
Witten's instanton is a conventional tunneling phenomenon.

In the next section, we ask
whether, in the presence of a
tachyon, the potential of \ref{bpot}\ is reduced, the barrier
perhaps even disappearing in accord with Horowitz's conjecture.

\section{Coupling a quasi-localized tachyon}

In Horowitz's analysis, the role of the black string is to provide
a situation in which the radius of the compact dimension shrinks to the string scale (so that winding tachyons appear) while keeping the curvature low and the fields slowly varying. Therefore, we will analyze the question of whether the black string catalyzes the Witten
process by asking whether there is an enhancement of the tunneling rate \ref{rate}\ in cases where the KK radius becomes comparable to the string scale.
The quasi-localized tachyon will be modelled by a scalar field with radial dependent mass term.
We start by considering the problem at large radius, where the $\alpha^\prime$
expansion is reliable.  Even here, in the core of the Witten bounce, the KK radius 
becomes small, and one might expect a negative mode corresponding to the existence of a lower energy configuration.

The action we consider is
\begin{equation}\label{action}
S=-\dfrac{1}{16\pi l_P^{n}}\int_V \sqrt{g}\mathcal{R}+\dfrac{1}{8\pi l_P^{n-1}}\int_{\partial V}[\mathcal{K}]d\Sigma-\frac{1}{2}\int_V \sqrt{g}\{g^{\mu\nu}\partial_{\mu}\phi \partial_{\nu}\phi+m^2(r)\phi^2\}
\end{equation}
The second term is the Gibbons-Hawking \cite{gibhawk}\ extrinsic curvature term.
The mass term is related to the radius of the fifth dimension and
becomes negative when it reaches the string scale.

A perturbative {\it closed string} state in flat space that winds $w$ times around a KK circle has a mass that depends on the radius of the circle 
\begin{equation}\label{mass}
m^2(r)=\bigl(\frac{w R}{\alpha^{\prime}}\bigr)^2+\frac{4}{\alpha^{\prime}}\bigl(N-c\bigr)
\end{equation} where $N$ is the left-movers level (which is equal to the right-movers levels when there are no momentum excitations) and $c$ is an order one model dependent number. E.g. in type II superstrings $c=1/2$ and in the bosonic string $c=1$.
As shown by Rohm \cite{rohm}\ imposing anti-periodic boundary conditions on spacetime fermions around the circle leads to a reversal of the usual GSO projection in odd winding sectors $w=2k+1$. Thus, denoting by $R_{eff}(r)$ the radius of the KK direction as a function of the radial coordinate $r$, if $R_{eff}(r)(2k+1)\leq 2\ell_s\sqrt{c}$ there will be a physical tachyonic state in the spectrum coming from this winding sector. Since we are interested in the first winding tachyon ($w=1$) the mass term is 
\begin{equation}
m^2(r)=\bigl( \frac{1}{\alpha^{\prime}}\bigr)^2g_{\theta\theta}-\frac{4 c}{\alpha^{\prime}}
\end{equation} where $d\chi=Rd\theta.$
The equations of motion for the tachyon field derived from \ref{action}\ are
\begin{equation}
\nabla^2\phi-m^2(r)\phi=0.
\end{equation} 
where the Laplacian is that appropriate to the background.

As we have seen in the previous section, the metric (at the turning point) of the classical configurations making up the potential barrier is given by \ref{anzats}
\begin{equation}
ds^2_{CD}=\frac{dr^2}{f_n(r)}+r^2d\Omega_n^2+f_n(r)R^2d\theta^2.
\end{equation}
Since we only expect to find an instability (connected with $\phi$) in the
region where $m^2(r)\leq 0$, namely when $g_{\theta\theta}\sim \ell_s,$ we can expand around $r=r_0$ by defining $r=r_0+\lambda^2/2R$ and get, to order $\lambda^2$ 
\begin{equation}\label{fa}
ds^2\approx  d\lambda^2+\lambda^2 d\theta^2 +r_0^2d\Omega_n^2
\end{equation} as was expected because $r_0$ is just a coordinate singularity. Defining Cartesian coordinates $\lambda^2=x^2+y^2$ the quadratic fluctuation operator for an S-wave is given by
\begin{equation}\label{lap}
(\partial_x^2+\partial_y^2)-\bigl( \frac{1}{\alpha^{\prime}}\bigr)^2(x^2+y^2)+\frac{4 c}{\alpha^{\prime}}
\end{equation} 
Comparing with the Schroedinger equation for a 2d harmonic oscillator
\begin{equation}
-\frac{\hbar^2(\partial_x^2+\partial_y^2)}{2m}\Psi+\frac{1}{2}m\Omega^2(x^2+y^2)\Psi=\bigl(N+1\bigr)\hbar\Omega\Psi,\quad N=0,1,\dots
\end{equation} we see that the quadratic fluctuation operator \ref{lap}\ has the spectrum of a shifted 2d harmonic oscillator with frequency $\Omega=\frac{2}{\alpha^{\prime}}.$ The ground state energy is given by
\begin{equation}\label{gsgs}
E_0=\dfrac{2-4c}{\alpha^{\prime}}.
\end{equation}
This ground state corresponds to a negative eigenmodes when $c>1/2$.

There are several observations to be made here. 
\begin{itemize}
 \item The question
 of existence of a negative eigenmode in the spectrum
involves a competition between order one numbers.
\item Since in all string models $c\leq 1$ a
negative eigenmode can only arise at most from the first winding tachyon.  This justifies
self-consistently our focus on this mode.
\item In type II superstrings, $c=1/2$, so there is no instability caused by
coupling the tachyon.  On the other hand, the stability is marginal, higher
order corrections in $\alpha^\prime$ could well turn the eigenvalue of this
mode negative.  The sign of the eigenvalue is likely to depend on the
details of the compactification. In that case, as explained below, it is likely that one can construct a similar solution with nearly the same action.
\item The eigenfunction for the lowest mode, for large $R$, has support only very near the bubble wall, $r-r_0<\frac{\alpha^{\prime}}{R}$:
\begin{equation}
\Psi_0= (\frac{1}{\pi\alpha^{\prime}})^{1\over 4}e^{-\frac{(r-r_0)R}{\alpha^{\prime}}}.
\end{equation}
\item  If in some model there is an instability for large $R$ we might still expect to find
a solution quite close to the Witten solution.  For example, if the tachyon potential includes a positive
quartic term\cite{dgk}, there will be a solution of slightly lower action with a small admixture of the unstable mode.
\end{itemize}

In conclusion, to first order in $\alpha^{\prime}$, coupling a quasi-localized tachyon seems to have a very small effect, if at all, on the potential barrier for tunneling into a BON. 

\section{Application to Horowitz's conjecture}

Recently, Horowitz \cite{horowitz} argued that black strings can catalyze Witten's decay process. Namely, if the KK vacuum suffers from the Witten instability at a non-perturbative and exponentially suppressed rate, {\it the excited state of a black ring}  in an asymptotically $R^{1,d}\times S^1$ space can enhance the decay rate dramatically. In this section we try to apply the lessons from the previous sections to this scenario.

\subsection{The black-ring solution}

The black string string-frame metric in $n+3$ spacetime dimensions is given by
\begin{equation}\label{bs}
\begin{split}
ds^2&=\dfrac{1}{H(r)}[-f(r)dt^2+R^2d\theta^2]+\dfrac{dr^2}{f(r)}+r^2d\Omega^2_n\\
B_{xt}&={\sinh 2\alpha\over 2H(r)}\quad,\quad e^{-2\Phi}=H(r)\qquad\rm{where} \\
f(r)&=1-\dfrac{r_H^{n-1}}{r^{n-1}}, \quad H(r)=1+\dfrac{r_H^{n-1}}{r^{n-1}}\sinh^2\alpha=\cosh^2\alpha-f(r)\sinh^2\alpha\\
\end{split}
\end{equation} 
The solution has a horizon at $r=r_H$ and a singularity at $r=0$. The winding number around the $\theta$ direction is proportional to 
\begin{equation}
W=(\frac{r_H}{\ell_s})^{n-1}\sinh 2\alpha.
\end{equation}  
This classical solution involves no fermionic fields and so exists also in the case where the fermions have anti-periodic boundary conditions around the KK circle, breaking supersymmetry.
The effective radius of the KK direction 
\begin{equation}\label{reff}
R^2_{eff}(r)\equiv g_{\theta\theta}(r)=\frac{R^2}{H(r)}
\end{equation} decreases with $r$ until it vanishes hitting the singularity at $r=0$.

One can choose the parameters $W\gg e^{2\alpha}\gg 1$ keeping the asymptotic value of the
KK radius large and the asymptotic string coupling small
\begin{equation}
 R\gg \ell_s\qquad\rm{and}\qquad g_s(\infty)\ll 1.
\end{equation}
In this regime, the black string has two important features. First, at the horizon $r=r_H$ the effective radius can be tuned to reach the string scale 
\begin{equation}\label{reffh}
R^2_{eff}(r_H)=\frac{R^2}{\cosh^2\alpha}\sim \alpha^{\prime}
\end{equation}
Secondly, the curvature and string coupling are small
all the way from the asymptotic region to the horizon:
\begin{equation}
\begin{split}
 \mathcal{R}_{r=r_H}&\sim\frac{1}{r_H^2}\sim(\frac{\sinh 2\alpha}{W})^{\frac{2}{n-1}}\cdot m_s^2\ll m_s^2\\
g_s(r_H)&=\frac{g_s(\infty)}{\cosh \alpha}\ll 1.
\end{split}
\end{equation} 

\subsection{A possible spacetime picture.}

In this region of parameter space we can approximate the physics outside and not too far away from the horizon with flat space formulas.
In principle, as we approach $r=0$ (keeping couplings small to allow the use of this flat space analysis) more and more winding tachyons enter into the spectrum according to \ref{mass}. To keep things simple we choose the first tachyon to appear just outside the horizon and concern ourselves only with the effects of that tachyon on an asymptotic observer.

\begin{figure}[h]
 \centering
 \includegraphics[width=8cm,height=6cm]{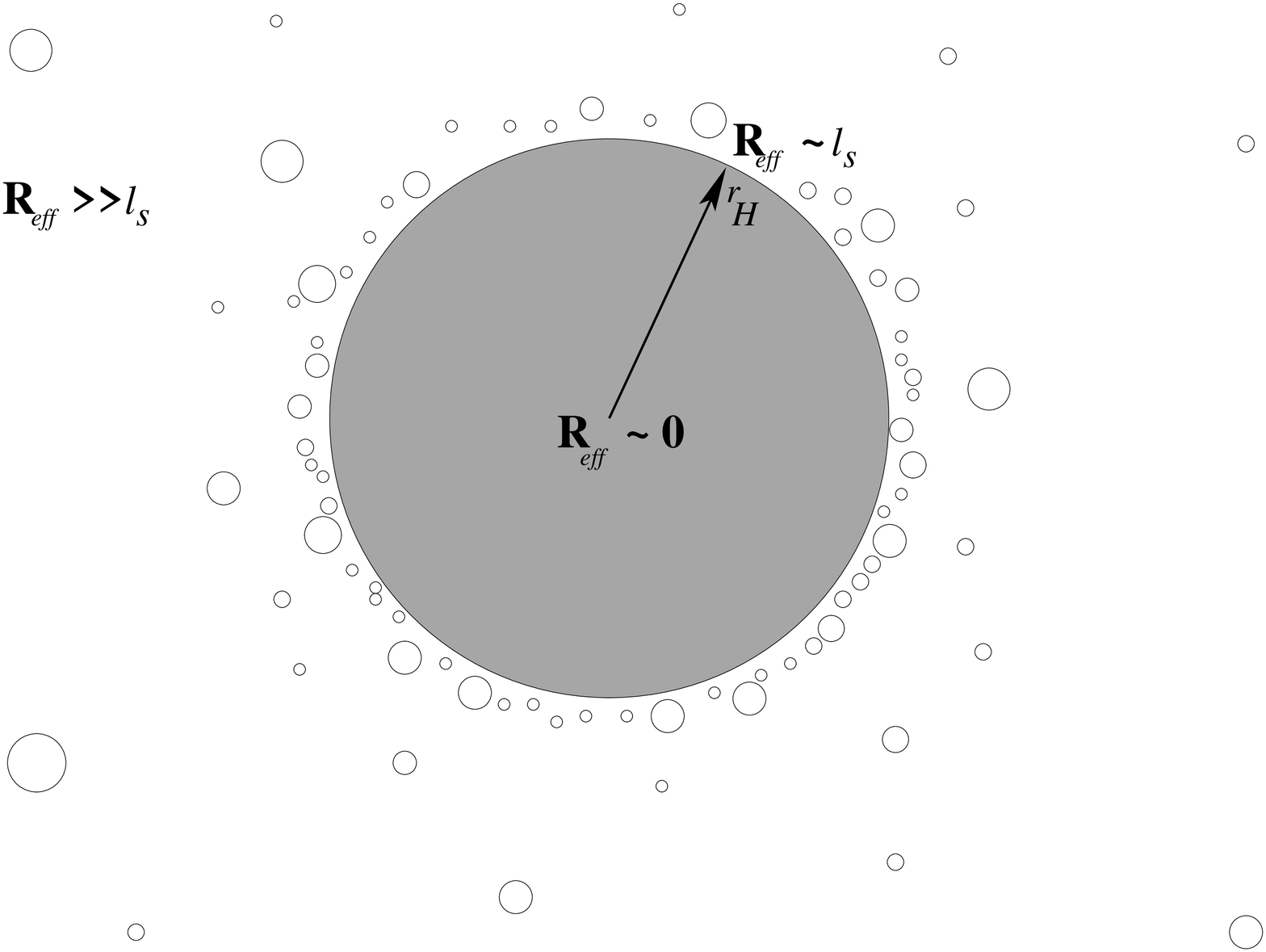}
\caption{{\bf A possible spacetime picture for the Horowitz process.}\newline \emph{This figure shows a section of the black ring. The curvature outside the horizon and far away is small so the space there is almost flat and suffers from the usual Witten instability. Near the horizon the effective radius reaches the string scale so winding tachyons appear, perhaps causing an enhancement of the decay rate into BONs (drawn as empty circles).}}
\end{figure}

The essential feature of the black ring is the geometrical fact that the effective radius of a circle shrinks as we move in radially in a fashion that allows to have a Scherk-Schwarz \cite{ss}\ winding tachyon just outside the horizon, keeping both the $g_s$ and $\alpha^{\prime}$ expansions under control. 
In such a set-up the space just outside the horizon is flat to a good approximation and as such subject to Witten's instability.
The fact that the rate of decay into a flat-space BON (eqn. \ref{rate}) becomes higher and higher as one moves in radially towards the black ring horizon lends some support to Horowitz's conjecture. 

At the horizon, where the first winding tachyon appears, the work of \cite{pinch}
makes it seem possible that the rate of decay becomes of order one. 
Our study of the tachyon
instability showing marginal stability in the presence of the tachyon indicates that this
is a real possibility.  However, one can not answer this question within
the range of validity of the $\alpha^{\prime}$ expansion.

\section{Conclusions}

In this note, we constructed a set of configurations which interpolate
between the KK vacuum and a bubble of nothing, and describe
the potential barrier.  For these configurations, we saw that, to leading order in $\alpha^\prime,$ the presence of the tachyon does not lead to a lowering of the barrier and an enhancement of the tunneling rate.
On the other hand, as the radius approaches the string scale, the action and the solution change so the analysis requires
modification.
If there are instabilities, it is necessary to understand the structure of the
tachyon effective action; for example, is the potential bounded
below?  In principle, this
is a problem of weak coupling, but for a general background, string techniques
do not presently exist to address it.  
Conceivably, in some cases, conformal field theory techniques
can be used to analyze this problem, along the lines of \cite{pinch}.  But for the
moment, we must admit various logical possibilities when $R\sim\ell_s$:
\begin{enumerate}
\item
The potential changes by an order one amount, but the amplitude \ref{rate}\ is still highly
suppressed at weak string coupling.
\item
The barrier disappears and the tunneling rate is unsuppressed.
\end{enumerate}
Energetically, for string scale compactifications, it is quite plausible that the
barrier to tunneling to nothing disappears.  In this case, Horowitz's observation
that topological defects can catalyze these processes is surely important.
It may indicate that even in regimes of moduli space where one would have thought
these states would be highly metastable, they are short-lived.  Conceivably, this
could be an argument that the world around us should have at least some approximate
supersymmetry.

Recently, string constructions
with all moduli stabilized, and which appear inherently perturbative, have
been put forward \cite{dewolfe}.  It appears possible to extend these constructions
to Scherk-Schwarz compactifications \cite{toappear}.  In such models, some of the questions
we have proposed may be sharper but it seems likely that also in this case higher $\alpha^{\prime}$ corrections will not be under control.

\subsection*{Acknowledgments}
It is a pleasure to thank Anthony Aguirre, Tom Banks, G. Horowitz, A.
Lawrence and E. Silverstein for useful discussions. 
This research is supported by DOE grant DE-FG03-92ER40689.

\end{document}